\documentclass[letterpaper,compsoc,twoside,final]{IEEEtran}
\usepackage{fixltx2e} 
\usepackage{cmap} 
\usepackage{}
\usepackage{ifthen}
\usepackage[T1]{fontenc}
\usepackage[utf8]{inputenc}
\usepackage{amsmath}
\usepackage{booktabs}
\usepackage{color}

\usepackage[font={small,it},labelfont=bf]{caption}
\usepackage{float}

\usepackage{scipy}
\makeatletter
\def\PY@reset{\let\PY@it=\relax \let\PY@bf=\relax%
    \let\PY@ul=\relax \let\PY@tc=\relax%
    \let\PY@bc=\relax \let\PY@ff=\relax}
\def\PY@tok#1{\csname PY@tok@#1\endcsname}
\def\PY@toks#1+{\ifx\relax#1\empty\else%
    \PY@tok{#1}\expandafter\PY@toks\fi}
\def\PY@do#1{\PY@bc{\PY@tc{\PY@ul{%
    \PY@it{\PY@bf{\PY@ff{#1}}}}}}}
\def\PY#1#2{\PY@reset\PY@toks#1+\relax+\PY@do{#2}}

\expandafter\def\csname PY@tok@gd\endcsname{\def\PY@tc##1{\textcolor[rgb]{0.63,0.00,0.00}{##1}}}
\expandafter\def\csname PY@tok@gu\endcsname{\let\PY@bf=\textbf\def\PY@tc##1{\textcolor[rgb]{0.50,0.00,0.50}{##1}}}
\expandafter\def\csname PY@tok@gt\endcsname{\def\PY@tc##1{\textcolor[rgb]{0.00,0.27,0.87}{##1}}}
\expandafter\def\csname PY@tok@gs\endcsname{\let\PY@bf=\textbf}
\expandafter\def\csname PY@tok@gr\endcsname{\def\PY@tc##1{\textcolor[rgb]{1.00,0.00,0.00}{##1}}}
\expandafter\def\csname PY@tok@cm\endcsname{\let\PY@it=\textit\def\PY@tc##1{\textcolor[rgb]{0.25,0.50,0.56}{##1}}}
\expandafter\def\csname PY@tok@vg\endcsname{\def\PY@tc##1{\textcolor[rgb]{0.73,0.38,0.84}{##1}}}
\expandafter\def\csname PY@tok@vi\endcsname{\def\PY@tc##1{\textcolor[rgb]{0.73,0.38,0.84}{##1}}}
\expandafter\def\csname PY@tok@vm\endcsname{\def\PY@tc##1{\textcolor[rgb]{0.73,0.38,0.84}{##1}}}
\expandafter\def\csname PY@tok@mh\endcsname{\def\PY@tc##1{\textcolor[rgb]{0.13,0.50,0.31}{##1}}}
\expandafter\def\csname PY@tok@cs\endcsname{\def\PY@tc##1{\textcolor[rgb]{0.25,0.50,0.56}{##1}}\def\PY@bc##1{\setlength{\fboxsep}{0pt}\colorbox[rgb]{1.00,0.94,0.94}{\strut ##1}}}
\expandafter\def\csname PY@tok@ge\endcsname{\let\PY@it=\textit}
\expandafter\def\csname PY@tok@vc\endcsname{\def\PY@tc##1{\textcolor[rgb]{0.73,0.38,0.84}{##1}}}
\expandafter\def\csname PY@tok@il\endcsname{\def\PY@tc##1{\textcolor[rgb]{0.13,0.50,0.31}{##1}}}
\expandafter\def\csname PY@tok@go\endcsname{\def\PY@tc##1{\textcolor[rgb]{0.20,0.20,0.20}{##1}}}
\expandafter\def\csname PY@tok@cp\endcsname{\def\PY@tc##1{\textcolor[rgb]{0.00,0.44,0.13}{##1}}}
\expandafter\def\csname PY@tok@gi\endcsname{\def\PY@tc##1{\textcolor[rgb]{0.00,0.63,0.00}{##1}}}
\expandafter\def\csname PY@tok@gh\endcsname{\let\PY@bf=\textbf\def\PY@tc##1{\textcolor[rgb]{0.00,0.00,0.50}{##1}}}
\expandafter\def\csname PY@tok@ni\endcsname{\let\PY@bf=\textbf\def\PY@tc##1{\textcolor[rgb]{0.84,0.33,0.22}{##1}}}
\expandafter\def\csname PY@tok@nl\endcsname{\let\PY@bf=\textbf\def\PY@tc##1{\textcolor[rgb]{0.00,0.13,0.44}{##1}}}
\expandafter\def\csname PY@tok@nn\endcsname{\let\PY@bf=\textbf\def\PY@tc##1{\textcolor[rgb]{0.05,0.52,0.71}{##1}}}
\expandafter\def\csname PY@tok@no\endcsname{\def\PY@tc##1{\textcolor[rgb]{0.38,0.68,0.84}{##1}}}
\expandafter\def\csname PY@tok@na\endcsname{\def\PY@tc##1{\textcolor[rgb]{0.25,0.44,0.63}{##1}}}
\expandafter\def\csname PY@tok@nb\endcsname{\def\PY@tc##1{\textcolor[rgb]{0.00,0.44,0.13}{##1}}}
\expandafter\def\csname PY@tok@nc\endcsname{\let\PY@bf=\textbf\def\PY@tc##1{\textcolor[rgb]{0.05,0.52,0.71}{##1}}}
\expandafter\def\csname PY@tok@nd\endcsname{\let\PY@bf=\textbf\def\PY@tc##1{\textcolor[rgb]{0.33,0.33,0.33}{##1}}}
\expandafter\def\csname PY@tok@ne\endcsname{\def\PY@tc##1{\textcolor[rgb]{0.00,0.44,0.13}{##1}}}
\expandafter\def\csname PY@tok@nf\endcsname{\def\PY@tc##1{\textcolor[rgb]{0.02,0.16,0.49}{##1}}}
\expandafter\def\csname PY@tok@si\endcsname{\let\PY@it=\textit\def\PY@tc##1{\textcolor[rgb]{0.44,0.63,0.82}{##1}}}
\expandafter\def\csname PY@tok@s2\endcsname{\def\PY@tc##1{\textcolor[rgb]{0.25,0.44,0.63}{##1}}}
\expandafter\def\csname PY@tok@nt\endcsname{\let\PY@bf=\textbf\def\PY@tc##1{\textcolor[rgb]{0.02,0.16,0.45}{##1}}}
\expandafter\def\csname PY@tok@nv\endcsname{\def\PY@tc##1{\textcolor[rgb]{0.73,0.38,0.84}{##1}}}
\expandafter\def\csname PY@tok@s1\endcsname{\def\PY@tc##1{\textcolor[rgb]{0.25,0.44,0.63}{##1}}}
\expandafter\def\csname PY@tok@dl\endcsname{\def\PY@tc##1{\textcolor[rgb]{0.25,0.44,0.63}{##1}}}
\expandafter\def\csname PY@tok@ch\endcsname{\let\PY@it=\textit\def\PY@tc##1{\textcolor[rgb]{0.25,0.50,0.56}{##1}}}
\expandafter\def\csname PY@tok@m\endcsname{\def\PY@tc##1{\textcolor[rgb]{0.13,0.50,0.31}{##1}}}
\expandafter\def\csname PY@tok@gp\endcsname{\let\PY@bf=\textbf\def\PY@tc##1{\textcolor[rgb]{0.78,0.36,0.04}{##1}}}
\expandafter\def\csname PY@tok@sh\endcsname{\def\PY@tc##1{\textcolor[rgb]{0.25,0.44,0.63}{##1}}}
\expandafter\def\csname PY@tok@ow\endcsname{\let\PY@bf=\textbf\def\PY@tc##1{\textcolor[rgb]{0.00,0.44,0.13}{##1}}}
\expandafter\def\csname PY@tok@sx\endcsname{\def\PY@tc##1{\textcolor[rgb]{0.78,0.36,0.04}{##1}}}
\expandafter\def\csname PY@tok@bp\endcsname{\def\PY@tc##1{\textcolor[rgb]{0.00,0.44,0.13}{##1}}}
\expandafter\def\csname PY@tok@c1\endcsname{\let\PY@it=\textit\def\PY@tc##1{\textcolor[rgb]{0.25,0.50,0.56}{##1}}}
\expandafter\def\csname PY@tok@fm\endcsname{\def\PY@tc##1{\textcolor[rgb]{0.02,0.16,0.49}{##1}}}
\expandafter\def\csname PY@tok@o\endcsname{\def\PY@tc##1{\textcolor[rgb]{0.40,0.40,0.40}{##1}}}
\expandafter\def\csname PY@tok@kc\endcsname{\let\PY@bf=\textbf\def\PY@tc##1{\textcolor[rgb]{0.00,0.44,0.13}{##1}}}
\expandafter\def\csname PY@tok@c\endcsname{\let\PY@it=\textit\def\PY@tc##1{\textcolor[rgb]{0.25,0.50,0.56}{##1}}}
\expandafter\def\csname PY@tok@mf\endcsname{\def\PY@tc##1{\textcolor[rgb]{0.13,0.50,0.31}{##1}}}
\expandafter\def\csname PY@tok@err\endcsname{\def\PY@bc##1{\setlength{\fboxsep}{0pt}\fcolorbox[rgb]{1.00,0.00,0.00}{1,1,1}{\strut ##1}}}
\expandafter\def\csname PY@tok@mb\endcsname{\def\PY@tc##1{\textcolor[rgb]{0.13,0.50,0.31}{##1}}}
\expandafter\def\csname PY@tok@ss\endcsname{\def\PY@tc##1{\textcolor[rgb]{0.32,0.47,0.09}{##1}}}
\expandafter\def\csname PY@tok@sr\endcsname{\def\PY@tc##1{\textcolor[rgb]{0.14,0.33,0.53}{##1}}}
\expandafter\def\csname PY@tok@mo\endcsname{\def\PY@tc##1{\textcolor[rgb]{0.13,0.50,0.31}{##1}}}
\expandafter\def\csname PY@tok@kd\endcsname{\let\PY@bf=\textbf\def\PY@tc##1{\textcolor[rgb]{0.00,0.44,0.13}{##1}}}
\expandafter\def\csname PY@tok@mi\endcsname{\def\PY@tc##1{\textcolor[rgb]{0.13,0.50,0.31}{##1}}}
\expandafter\def\csname PY@tok@kn\endcsname{\let\PY@bf=\textbf\def\PY@tc##1{\textcolor[rgb]{0.00,0.44,0.13}{##1}}}
\expandafter\def\csname PY@tok@cpf\endcsname{\let\PY@it=\textit\def\PY@tc##1{\textcolor[rgb]{0.25,0.50,0.56}{##1}}}
\expandafter\def\csname PY@tok@kr\endcsname{\let\PY@bf=\textbf\def\PY@tc##1{\textcolor[rgb]{0.00,0.44,0.13}{##1}}}
\expandafter\def\csname PY@tok@s\endcsname{\def\PY@tc##1{\textcolor[rgb]{0.25,0.44,0.63}{##1}}}
\expandafter\def\csname PY@tok@kp\endcsname{\def\PY@tc##1{\textcolor[rgb]{0.00,0.44,0.13}{##1}}}
\expandafter\def\csname PY@tok@w\endcsname{\def\PY@tc##1{\textcolor[rgb]{0.73,0.73,0.73}{##1}}}
\expandafter\def\csname PY@tok@kt\endcsname{\def\PY@tc##1{\textcolor[rgb]{0.56,0.13,0.00}{##1}}}
\expandafter\def\csname PY@tok@sc\endcsname{\def\PY@tc##1{\textcolor[rgb]{0.25,0.44,0.63}{##1}}}
\expandafter\def\csname PY@tok@sb\endcsname{\def\PY@tc##1{\textcolor[rgb]{0.25,0.44,0.63}{##1}}}
\expandafter\def\csname PY@tok@sa\endcsname{\def\PY@tc##1{\textcolor[rgb]{0.25,0.44,0.63}{##1}}}
\expandafter\def\csname PY@tok@k\endcsname{\let\PY@bf=\textbf\def\PY@tc##1{\textcolor[rgb]{0.00,0.44,0.13}{##1}}}
\expandafter\def\csname PY@tok@se\endcsname{\let\PY@bf=\textbf\def\PY@tc##1{\textcolor[rgb]{0.25,0.44,0.63}{##1}}}
\expandafter\def\csname PY@tok@sd\endcsname{\let\PY@it=\textit\def\PY@tc##1{\textcolor[rgb]{0.25,0.44,0.63}{##1}}}

\def\PYZus{\char`\_}
\def\PYZob{\char`\{}
\def\PYZcb{\char`\}}
\def\PYZca{\char`\^}

\def\PYZlt{\char`\<}
\def\PYZgt{\char`\>}

\def\PYZhy{\char`\-}
\def\PYZsq{\char`\'}
\def\PYZdq{\char`\"}

\makeatother

\providecommand*\DUrolestring[1]{\textit{#1}}

\providecommand*{\DUrole}[2]{%
  \ifcsname DUrole#1\endcsname%
    \csname DUrole#1\endcsname{#2}%
  \else
    \ifcsname docutilsrole#1\endcsname%
      \csname docutilsrole#1\endcsname{#2}%
    \else%
      #2%
    \fi%
  \fi%
}

\providecommand*{\DUroletitlereference}[1]{\textsl{#1}}

\ifthenelse{\isundefined{\hypersetup}}{
  \usepackage[colorlinks=true,linkcolor=blue,urlcolor=blue]{hyperref}
  \urlstyle{same} 
}{}

\begin{document}
\newcounter{footnotecounter}\title{MatchPy: A Pattern Matching Library}\author{Manuel Krebber$^{\setcounter{footnotecounter}{3}\fnsymbol{footnotecounter}\setcounter{footnotecounter}{1}\fnsymbol{footnotecounter}}$%
          \setcounter{footnotecounter}{1}\thanks{\fnsymbol{footnotecounter} %
          Corresponding author: \protect\href{mailto:manuel.krebber@rwth-aachen.de}{manuel.krebber@rwth-aachen.de}}\setcounter{footnotecounter}{3}\thanks{\fnsymbol{footnotecounter} RWTH Aachen University, AICES, HPAC Group}, Henrik Bartels$^{\setcounter{footnotecounter}{3}\fnsymbol{footnotecounter}}$, Paolo Bientinesi$^{\setcounter{footnotecounter}{3}\fnsymbol{footnotecounter}}$\thanks{%

          \noindent%
          Copyright\,\copyright\,2017 Manuel Krebber et al. This is an open-access article distributed under the terms of the Creative Commons Attribution License, which permits unrestricted use, distribution, and reproduction in any medium, provided the original author and source are credited.%
        }}\maketitle
          \renewcommand{\leftmark}{PROC. OF THE 15th PYTHON IN SCIENCE CONF. (SCIPY 2017)}
          \renewcommand{\rightmark}{MATCHPY: A PATTERN MATCHING LIBRARY}

\InputIfFileExists{page_numbers.tex}{}{}
\newcommand*{\docutilsroleref}{\ref}
\newcommand*{\docutilsrolelabel}{\label}
\providecommand*\DUrolecite[1]{\cite{#1}}

\providecommand*\DUrolebuiltin[1]{\PY{n+nb}{#1}}
\providecommand*\DUrolepython[1]{\footnotesize{#1}}
\renewcommand\DUrolestring[1]{\PY{l+s+s1}{#1}}

\begin{abstract}Pattern matching is a powerful tool for symbolic computations, based on the well-defined theory of term rewriting systems.
Application domains include algebraic expressions, abstract syntax trees, and XML and JSON data.
Unfortunately, no lightweight implementation of pattern matching as general and flexible as Mathematica exists for Python \DUrole{cite}{Mathics,MacroPy,patterns,PyPatt}.
Therefore, we created the open source module \href{https://github.com/HPAC/matchpy}{MatchPy} which offers similar pattern matching functionality in Python using
a novel algorithm which finds matches for large pattern sets more efficiently by exploiting similarities between patterns.\end{abstract}

\subsection{Introduction%
  \label{introduction}%
}

Pattern matching is a powerful tool which is part of many functional programming languages as well as computer algebra systems such as Mathematica.
It is useful for many applications including symbolic computation, term simplification, term rewriting systems, automated theorem proving, and model checking.
In this paper, we present a Python-based pattern matching library and its underlying algorithms.

The goal of pattern matching is to find a match substitution given a subject term and a pattern which is a term with placeholders \DUrole{cite}{Baader1998}.
The substitution maps placeholders in the pattern to replacement terms.
A match is a substitution that can be applied to the pattern yielding the original subject.
As an example consider the subject $f(a)$ and the pattern $f(x)$ where $x$ is a placeholder.
Then the substitution $\sigma = \{ x \mapsto a \}$ is a match because $\sigma(f(x)) = f(a)$.
This form of pattern matching without any special properties of function symbols is called syntactic matching.
For syntactic patterns, the match is unique if it exists.

Among the existing systems, Mathematica \DUrole{cite}{Mathematica} arguably offers the most expressive pattern matching.
Its pattern matching offers similar expressiveness as Python's regular expressions, but for symbolic tree structures instead of strings.
While pattern matching can handle nested expressions up to arbitrary depth, regular expressions cannot properly handle such nesting.
Patterns are used widely in Mathematica, e.g. in function definitions or for manipulating terms.
It is possible to define custom function symbols which can be associative and/or commutative.
Mathematica also offers sequence variables which can match a sequence of terms instead of a single one.
These are especially useful when working with variadic function symbols.
Mathics \DUrole{cite}{Mathics} is an open source computer algebra system written in Python that aims to replicate the syntax and functionality of Mathematica.

To our knowledge, no existing work covers pattern matching with function symbols which are either commutative or associative but not both at the same time.
However, there are functions with those properties, e.g. matrix multiplication or arithmetic mean.
Most of the existing pattern matching libraries for Python only support syntactic patterns.
Associativity, commutativity and sequence variables make multiple distinct matches possible for a single pattern.
In addition, pattern matching with either associativity or commutativity is NP-complete in both cases \DUrole{cite}{Benanav1987}.
While the pattern matching in SymPy \DUrole{cite}{SymPy} can work with associative/commutative functions, it is limited to finding a single match.
Nonetheless, for some applications it is interesting to find all possible matches for a pattern,
e.g. because matches need to be processed further recursively to solve an optimization problem.
Furthermore, SymPy does not support sequence variables and is limited to a predefined set of mathematical operations.

In many applications, a fixed set of patterns is matched repeatedly against different subjects.
The simultaneous matching of multiple patterns is called many-to-one matching, as opposed to one-to-one matching which denotes matching with a single pattern.
Many-to-one matching can gain a significant speed increase compared to one-to-one matching by exploiting similarities between patterns.
This has already been the subject of research for both syntactic \DUrole{cite}{Christian1993,Graef1991,Nedjah1997}
and associative-commutative pattern matching \DUrole{cite}{Kounalis1991,Bachmair1993,Lugiez1994,Bachmair1995,Eker1995,Kirchner2001}, but not with the full feature set described above.
Discrimination nets \DUrole{cite}{Bundy1984} are the state-of-the-art solution for many-to-one matching.
Our goal is to generalize this approach to support all the aforementioned features.

In this paper, we present the open-source library for Python \href{https://github.com/HPAC/matchpy}{MatchPy} which provides pattern matching with sequence variables and associative/commutative function symbols.
In addition to standard one-to-one matching, MatchPy also includes an efficient many-to-one matching algorithm that uses generalized discrimination nets.
First, we give an overview of what MatchPy can be used for.
Secondly, we explain some of the challenges arising from the non-syntactic pattern matching features and how we solve them.
Then we give an overview of how many-to-one matching is realized and optimized in MatchPy.
Next, we present our experiments where we observed significant speedups of the many-to-one matching over one-to-one matching.
Finally, we draw some conclusions from the experiments and propose future work on MatchPy.

\subsection{Usage Overview%
  \label{usage-overview}%
}

MatchPy can be installed using \texttt{pip} and all necessary classes can be imported from the toplevel module \texttt{matchpy}.
Expressions in MatchPy consist of constant symbols and operations.
For patterns, wildcards can also be used as placeholders.
We use \href{https://reference.wolfram.com/language/guide/Patterns.html}{Mathematica's notation} for wildcards,
i.e. we append underscores to wildcard names to distinguish them from symbols.

MatchPy can be used with native Python types such as \texttt{list} and \texttt{int}.
The following is an example of how the subject \texttt{{[}0, 1{]}} can be matched against the pattern \texttt{{[}x\_, 1{]}}.
The expected match here is the replacement \texttt{0} for \texttt{x\_}.
We use \DUroletitlereference{next} because we only want to use the first (and in this case only) match of the pattern:\vspace{1mm}
\begin{Verbatim}[commandchars=\\\{\},fontsize=\footnotesize]
\PY{g+gp}{\PYZgt{}\PYZgt{}\PYZgt{} }\PY{n}{x\PYZus{}} \PY{o}{=} \PY{n}{Wildcard}\PY{o}{.}\PY{n}{dot}\PY{p}{(}\PY{l+s+s1}{\PYZsq{}}\PY{l+s+s1}{x}\PY{l+s+s1}{\PYZsq{}}\PY{p}{)}
\PY{g+gp}{\PYZgt{}\PYZgt{}\PYZgt{} }\PY{n+nb}{next}\PY{p}{(}\PY{n}{match}\PY{p}{(}\PY{p}{[}\PY{l+m+mi}{0}\PY{p}{,} \PY{l+m+mi}{1}\PY{p}{]}\PY{p}{,} \PY{n}{Pattern}\PY{p}{(}\PY{p}{[}\PY{n}{x\PYZus{}}\PY{p}{,} \PY{l+m+mi}{1}\PY{p}{]}\PY{p}{)}\PY{p}{)}\PY{p}{)}
\PY{g+go}{\PYZob{}\PYZsq{}x\PYZsq{}: 0\PYZcb{}}
\end{Verbatim}
\vspace{1mm}
In addition to regular (dot) variables, MatchPy also supports sequence wildcards.
They can match a sequence of arguments and we denote them with two or three trailing underscores for plus and star wildcards, respectively.
Star wildcards can match an empty sequence, while plus wildcards require at least one argument to match.
The terminology is borrowed from regular expressions where \DUroletitlereference{*}, \DUroletitlereference{+} and \DUroletitlereference{.} are used for similar concepts.\vspace{1mm}
\begin{Verbatim}[commandchars=\\\{\},fontsize=\footnotesize]
\PY{g+gp}{\PYZgt{}\PYZgt{}\PYZgt{} }\PY{n}{y\PYZus{}\PYZus{}\PYZus{}} \PY{o}{=} \PY{n}{Wildcard}\PY{o}{.}\PY{n}{star}\PY{p}{(}\PY{l+s+s1}{\PYZsq{}}\PY{l+s+s1}{y}\PY{l+s+s1}{\PYZsq{}}\PY{p}{)}
\PY{g+gp}{\PYZgt{}\PYZgt{}\PYZgt{} }\PY{n+nb}{next}\PY{p}{(}\PY{n}{match}\PY{p}{(}\PY{p}{[}\PY{l+m+mi}{1}\PY{p}{,} \PY{l+m+mi}{2}\PY{p}{,} \PY{l+m+mi}{3}\PY{p}{]}\PY{p}{,} \PY{n}{Pattern}\PY{p}{(}\PY{p}{[}\PY{n}{x\PYZus{}}\PY{p}{,} \PY{n}{y\PYZus{}\PYZus{}\PYZus{}}\PY{p}{]}\PY{p}{)}\PY{p}{)}\PY{p}{)}
\PY{g+go}{\PYZob{}\PYZsq{}x\PYZsq{}: 1, \PYZsq{}y\PYZsq{}: (2, 3)\PYZcb{}}
\end{Verbatim}
\vspace{1mm}
In the following, we omit the definition of new variables as they can be done in the same way.
In addition to native types, one can also define custom operations by creating a subclass of the \texttt{Operation} class:\vspace{1mm}
\begin{Verbatim}[commandchars=\\\{\},fontsize=\footnotesize]
\PY{k}{class} \PY{n+nc}{MyOp}\PY{p}{(}\PY{n}{Operation}\PY{p}{)}\PY{p}{:}
  \PY{n}{name} \PY{o}{=} \PY{l+s+s1}{\PYZsq{}}\PY{l+s+s1}{MyOp}\PY{l+s+s1}{\PYZsq{}}
  \PY{n}{arity} \PY{o}{=} \PY{n}{Arity}\PY{o}{.}\PY{n}{variadic}
  \PY{n}{associative} \PY{o}{=} \PY{n+nb+bp}{True}
  \PY{n}{commutative} \PY{o}{=} \PY{n+nb+bp}{True}
\end{Verbatim}
\vspace{1mm}
The name is a required attribute, while the others are optional and influence the behavior of the operations.
By default, operations are variadic and neither commutative nor associative.
Nested associative operations have to be variadic and are automatically flattened.
Furthermore, regular variables behave similar to sequence variables as arguments of associative functions,
because the associativity allows arbitrary parenthesization of arguments:\vspace{1mm}
\begin{Verbatim}[commandchars=\\\{\},fontsize=\footnotesize]
\PY{g+gp}{\PYZgt{}\PYZgt{}\PYZgt{} }\PY{n+nb}{next}\PY{p}{(}\PY{n}{match}\PY{p}{(}\PY{n}{MyOp}\PY{p}{(}\PY{l+m+mi}{0}\PY{p}{,} \PY{l+m+mi}{1}\PY{p}{,} \PY{l+m+mi}{2}\PY{p}{)}\PY{p}{,} \PY{n}{Pattern}\PY{p}{(}\PY{n}{MyOp}\PY{p}{(}\PY{n}{x\PYZus{}}\PY{p}{,} \PY{l+m+mi}{2}\PY{p}{)}\PY{p}{)}\PY{p}{)}\PY{p}{)}
\PY{g+go}{\PYZob{}\PYZsq{}x\PYZsq{}: MyOp(0, 1)\PYZcb{}}
\end{Verbatim}
\vspace{1mm}
The argument of commutative operations are automatically sorted.
Note that patterns with commutative operations can have multiple matches, because their arguments can be reordered arbitrarily.\vspace{1mm}
\begin{Verbatim}[commandchars=\\\{\},fontsize=\footnotesize]
\PY{g+gp}{\PYZgt{}\PYZgt{}\PYZgt{} }\PY{n+nb}{list}\PY{p}{(}\PY{n}{match}\PY{p}{(}\PY{n}{MyOp}\PY{p}{(}\PY{l+m+mi}{1}\PY{p}{,} \PY{l+m+mi}{2}\PY{p}{)}\PY{p}{,} \PY{n}{Pattern}\PY{p}{(}\PY{n}{MyOp}\PY{p}{(}\PY{n}{x\PYZus{}}\PY{p}{,} \PY{n}{z\PYZus{}}\PY{p}{)}\PY{p}{)}\PY{p}{)}\PY{p}{)}
\PY{g+go}{[\PYZob{}\PYZsq{}x\PYZsq{}: 2, \PYZsq{}z\PYZsq{}: 1\PYZcb{}, \PYZob{}\PYZsq{}x\PYZsq{}: 1, \PYZsq{}z\PYZsq{}: 2\PYZcb{}]}
\end{Verbatim}
\vspace{1mm}
We can use the \texttt{CustomConstraint} class to create a constraint that checks whether \texttt{a} is smaller than \texttt{b}:\vspace{1mm}
\begin{Verbatim}[commandchars=\\\{\},fontsize=\footnotesize]
\PY{n}{a\PYZus{}lt\PYZus{}b} \PY{o}{=} \PY{n}{CustomConstraint}\PY{p}{(}\PY{k}{lambda} \PY{n}{a}\PY{p}{,} \PY{n}{b}\PY{p}{:} \PY{n}{a} \PY{o}{\PYZlt{}} \PY{n}{b}\PY{p}{)}
\end{Verbatim}
\vspace{1mm}
The lambda function gets called with the variable substitutions based on their name.
The order of arguments is not important and it is possible to only use a subset of the variables in the pattern.
With this constraint we can define a replacement rule that basically describes bubble sort:\vspace{1mm}
\begin{Verbatim}[commandchars=\\\{\},fontsize=\footnotesize]
\PY{g+gp}{\PYZgt{}\PYZgt{}\PYZgt{} }\PY{n}{pattern} \PY{o}{=} \PY{n}{Pattern}\PY{p}{(}\PY{p}{[}\PY{n}{h\PYZus{}\PYZus{}\PYZus{}}\PY{p}{,} \PY{n}{b\PYZus{}}\PY{p}{,} \PY{n}{a\PYZus{}}\PY{p}{,} \PY{n}{t\PYZus{}\PYZus{}\PYZus{}}\PY{p}{]}\PY{p}{,} \PY{n}{a\PYZus{}lt\PYZus{}b}\PY{p}{)}
\PY{g+gp}{\PYZgt{}\PYZgt{}\PYZgt{} }\PY{n}{rule} \PY{o}{=} \PY{n}{ReplacementRule}\PY{p}{(}\PY{n}{pattern}\PY{p}{,}
\PY{g+go}{                lambda a, b, h, t: [*h, a, b, *t])}
\end{Verbatim}
\vspace{1mm}
The replacement function gets called with all matched variables as keyword arguments and needs to return the replacement.
This replacement rule can be used to sort a list when applied repeatedly with \texttt{replace\_all}:\vspace{1mm}
\begin{Verbatim}[commandchars=\\\{\},fontsize=\footnotesize]
\PY{g+gp}{\PYZgt{}\PYZgt{}\PYZgt{} }\PY{n}{replace\PYZus{}all}\PY{p}{(}\PY{p}{[}\PY{l+m+mi}{1}\PY{p}{,} \PY{l+m+mi}{4}\PY{p}{,} \PY{l+m+mi}{3}\PY{p}{,} \PY{l+m+mi}{2}\PY{p}{]}\PY{p}{,} \PY{p}{[}\PY{n}{rule}\PY{p}{]}\PY{p}{)}
\PY{g+go}{[1, 2, 3, 4]}
\end{Verbatim}
\vspace{1mm}
Sequence variables can also be used to match subsequences that match a constraint.
For example, we can use the this feature to find all subsequences of integers that sum up to 5.
In the following example, we use anonymous wildcards which have no name and are hence not part of the match substitution:\vspace{1mm}
\begin{Verbatim}[commandchars=\\\{\},fontsize=\footnotesize]
\PY{g+gp}{\PYZgt{}\PYZgt{}\PYZgt{} }\PY{n}{x\PYZus{}sums\PYZus{}to\PYZus{}5} \PY{o}{=} \PY{n}{CustomConstraint}\PY{p}{(}
\PY{g+gp}{... }                        \PY{k}{lambda} \PY{n}{x}\PY{p}{:} \PY{n+nb}{sum}\PY{p}{(}\PY{n}{x}\PY{p}{)} \PY{o}{==} \PY{l+m+mi}{5}\PY{p}{)}
\PY{g+gp}{\PYZgt{}\PYZgt{}\PYZgt{} }\PY{n}{pattern} \PY{o}{=} \PY{n}{Pattern}\PY{p}{(}\PY{p}{[}\PY{n}{\PYZus{}\PYZus{}\PYZus{}}\PY{p}{,} \PY{n}{x\PYZus{}\PYZus{}}\PY{p}{,} \PY{n}{\PYZus{}\PYZus{}\PYZus{}}\PY{p}{]}\PY{p}{,} \PY{n}{x\PYZus{}sums\PYZus{}to\PYZus{}5}\PY{p}{)}
\PY{g+gp}{\PYZgt{}\PYZgt{}\PYZgt{} }\PY{n+nb}{list}\PY{p}{(}\PY{n}{match}\PY{p}{(}\PY{p}{[}\PY{l+m+mi}{1}\PY{p}{,} \PY{l+m+mi}{2}\PY{p}{,} \PY{l+m+mi}{3}\PY{p}{,} \PY{l+m+mi}{1}\PY{p}{,} \PY{l+m+mi}{1}\PY{p}{,} \PY{l+m+mi}{2}\PY{p}{]}\PY{p}{,} \PY{n}{pattern}\PY{p}{)}\PY{p}{)}
\PY{g+go}{[\PYZob{}\PYZsq{}x\PYZsq{}: (2, 3)\PYZcb{}, \PYZob{}\PYZsq{}x\PYZsq{}: (3, 1, 1)\PYZcb{}]}
\end{Verbatim}
\vspace{1mm}
More examples can be found in MatchPy's documentation \DUrole{cite}{MatchPyDoc}.

\subsubsection{Application Example: Finding matches for a BLAS kernel%
  \label{application-example-finding-matches-for-a-blas-kernel}%
}


\begin{table}
    \centering
    \renewcommand{\arraystretch}{1.2}
    \begin{tabular}{l c c p{1.5cm}}
        \toprule
        \textbf{Operation} & \textbf{Symbol} & \textbf{Arity} & \textbf{Properties} \\
        \midrule
        Multiplication & $\times$ & variadic & associative \\
        Addition & $+$ & variadic & associative,\newline commutative \\
        Transposition & ${}^T$ & unary & \\
        Inversion & ${}^{-1}$ & unary & \\
        Inversion and Transposition & ${}^{-T}$ & unary & \\
        \bottomrule
    \end{tabular}
    \caption{Linear Algebra Operations}
\label{tbl:laop}
\end{table}

BLAS is a collection of optimized routines that can compute specific linear algebra operations efficiently \DUrole{cite}{Lawson1979,Dongarra1988,Dongarra1990}.
As an example, assume we want to match all subexpressions of a linear algebra expression which can be computed by the \href{https://software.intel.com/en-us/node/468494}{?TRMM} BLAS routine.
These have the form $\alpha \times op(A)  \times B$ or $\alpha  \times B  \times op(A)$ where
$op(A)$ is either the identity function or transposition, and $A$ is a triangular matrix.
For this example, we leave out all variants where $\alpha \neq 1$.

In order to model the linear algebra expressions, we use the operations shown in Table \DUrole{ref}{tbl:laop}.
In addition, we have special symbol subclasses for scalars, vectors and matrices.
Matrices also have a set of properties, e.g. they can be triangular, symmetric, square, etc.
For those patterns we also use a special kind of dot variable which is restricted to only match a specific kind of symbol.
Finally, we construct the patterns using sequence variables to capture the remaining operands of the multiplication:\vspace{1mm}
\begin{Verbatim}[commandchars=\\\{\},fontsize=\footnotesize]
\PY{n}{A\PYZus{}} \PY{o}{=} \PY{n}{Wildcard}\PY{o}{.}\PY{n}{symbol}\PY{p}{(}\PY{l+s+s1}{\PYZsq{}}\PY{l+s+s1}{A}\PY{l+s+s1}{\PYZsq{}}\PY{p}{,} \PY{n}{Matrix}\PY{p}{)}
\PY{n}{B\PYZus{}} \PY{o}{=} \PY{n}{Wildcard}\PY{o}{.}\PY{n}{symbol}\PY{p}{(}\PY{l+s+s1}{\PYZsq{}}\PY{l+s+s1}{B}\PY{l+s+s1}{\PYZsq{}}\PY{p}{,} \PY{n}{Matrix}\PY{p}{)}
\PY{n}{A\PYZus{}is\PYZus{}triangular} \PY{o}{=} \PY{n}{CustomConstraint}\PY{p}{(}
  \PY{k}{lambda} \PY{n}{A}\PY{p}{:} \PY{l+s+s1}{\PYZsq{}}\PY{l+s+s1}{triangular}\PY{l+s+s1}{\PYZsq{}} \PY{o+ow}{in} \PY{n}{A}\PY{o}{.}\PY{n}{properties}\PY{p}{)}

\PY{n}{trmm\PYZus{}patterns} \PY{o}{=} \PY{p}{[}
  \PY{n}{Pattern}\PY{p}{(}\PY{n}{Times}\PY{p}{(}\PY{n}{h\PYZus{}\PYZus{}\PYZus{}}\PY{p}{,} \PY{n}{A\PYZus{}}\PY{p}{,} \PY{n}{B\PYZus{}}\PY{p}{,} \PY{n}{t\PYZus{}\PYZus{}\PYZus{}}\PY{p}{)}\PY{p}{,}
    \PY{n}{A\PYZus{}is\PYZus{}triangular}\PY{p}{)}\PY{p}{,}
  \PY{n}{Pattern}\PY{p}{(}\PY{n}{Times}\PY{p}{(}\PY{n}{h\PYZus{}\PYZus{}\PYZus{}}\PY{p}{,} \PY{n}{Transpose}\PY{p}{(}\PY{n}{A\PYZus{}}\PY{p}{)}\PY{p}{,} \PY{n}{B\PYZus{}}\PY{p}{,} \PY{n}{t\PYZus{}\PYZus{}\PYZus{}}\PY{p}{)}\PY{p}{,}
    \PY{n}{A\PYZus{}is\PYZus{}triangular}\PY{p}{)}\PY{p}{,}
  \PY{n}{Pattern}\PY{p}{(}\PY{n}{Times}\PY{p}{(}\PY{n}{h\PYZus{}\PYZus{}\PYZus{}}\PY{p}{,} \PY{n}{B\PYZus{}}\PY{p}{,} \PY{n}{A\PYZus{}}\PY{p}{,} \PY{n}{t\PYZus{}\PYZus{}\PYZus{}}\PY{p}{)}\PY{p}{,}
    \PY{n}{A\PYZus{}is\PYZus{}triangular}\PY{p}{)}\PY{p}{,}
  \PY{n}{Pattern}\PY{p}{(}\PY{n}{Times}\PY{p}{(}\PY{n}{h\PYZus{}\PYZus{}\PYZus{}}\PY{p}{,} \PY{n}{B\PYZus{}}\PY{p}{,} \PY{n}{Transpose}\PY{p}{(}\PY{n}{A\PYZus{}}\PY{p}{)}\PY{p}{,} \PY{n}{t\PYZus{}\PYZus{}\PYZus{}}\PY{p}{)}\PY{p}{,}
    \PY{n}{A\PYZus{}is\PYZus{}triangular}\PY{p}{)}\PY{p}{,}
\PY{p}{]}
\end{Verbatim}
\vspace{1mm}
With these patterns, we can find all matches for the \href{https://software.intel.com/en-us/node/468494}{?TRMM} routine within a product.
In this example, \texttt{M1}, \texttt{M2} and \texttt{M3} are matrices, but only \texttt{M3} is triangular:\vspace{1mm}
\begin{Verbatim}[commandchars=\\\{\},fontsize=\footnotesize]
\PY{g+gp}{\PYZgt{}\PYZgt{}\PYZgt{} }\PY{n}{expr} \PY{o}{=} \PY{n}{Times}\PY{p}{(}\PY{n}{Transpose}\PY{p}{(}\PY{n}{M3}\PY{p}{)}\PY{p}{,} \PY{n}{M1}\PY{p}{,} \PY{n}{M3}\PY{p}{,} \PY{n}{M2}\PY{p}{)}
\PY{g+gp}{\PYZgt{}\PYZgt{}\PYZgt{} }\PY{k}{for} \PY{n}{i}\PY{p}{,} \PY{n}{pattern} \PY{o+ow}{in} \PY{n+nb}{enumerate}\PY{p}{(}\PY{n}{trmm\PYZus{}patterns}\PY{p}{)}\PY{p}{:}
\PY{g+gp}{... }  \PY{k}{for} \PY{n}{substitution} \PY{o+ow}{in} \PY{n}{match}\PY{p}{(}\PY{n}{expr}\PY{p}{,} \PY{n}{pattern}\PY{p}{)}\PY{p}{:}
\PY{g+gp}{... }    \PY{k}{print}\PY{p}{(}\PY{l+s+s1}{\PYZsq{}}\PY{l+s+s1}{\PYZob{}\PYZcb{} with \PYZob{}\PYZcb{}}\PY{l+s+s1}{\PYZsq{}}\PY{o}{.}\PY{n}{format}\PY{p}{(}\PY{n}{i}\PY{p}{,} \PY{n}{substitution}\PY{p}{)}\PY{p}{)}
\PY{g+go}{0 with \PYZob{}A \PYZhy{}\PYZgt{} M3, B \PYZhy{}\PYZgt{} M2, t \PYZhy{}\PYZgt{} (), h \PYZhy{}\PYZgt{} ((M3)\PYZca{}T, M1)\PYZcb{}}
\PY{g+go}{1 with \PYZob{}A \PYZhy{}\PYZgt{} M3, B \PYZhy{}\PYZgt{} M1, t \PYZhy{}\PYZgt{} (M3, M2), h \PYZhy{}\PYZgt{} ()\PYZcb{}}
\PY{g+go}{2 with \PYZob{}A \PYZhy{}\PYZgt{} M3, B \PYZhy{}\PYZgt{} M1, t \PYZhy{}\PYZgt{} (M2), h \PYZhy{}\PYZgt{} ((M3)\PYZca{}T)\PYZcb{}}
\end{Verbatim}
\vspace{1mm}
As can be seen in the output, a total of three matches are found.

\subsection{Design Considerations%
  \label{design-considerations}%
}

There are plenty of implementations of syntactic matching and the algorithms are well known.
Implementing pattern matching for MatchPy poses some challenges such as associativity and commutativity.

\subsubsection{Associativity/Sequence variables%
  \label{associativity-sequence-variables}%
}

Associativity enables arbitrary grouping of arguments for matching:
For example, \texttt{1 + a + b} matches \texttt{1 + x\_} with $\{ x \mapsto a + b \}$ because we can group the arguments as \texttt{1 + (a + b)}.
Basically, when regular variables are arguments of an associative function, they behave like sequence variables.
Both can result in multiple distinct matches for a single pattern.
In contrast, for syntactic patterns there is always at most one match.
This means that the matching algorithm needs to be non-deterministic to explore all potential matches for associative terms or terms with sequence variables.
We employ backtracking with the help of Python generators to enable this.
Associative matching is NP-complete \DUrole{cite}{Benanav1987}.

\subsubsection{Commutativity%
  \label{commutativity}%
}

Matching commutative terms is difficult because matches need to be found independent of the argument order.
Commutative matching has been shown to be NP-complete, too \DUrole{cite}{Benanav1987}.
It is possible to find all matches by matching all permutations of the subjects arguments against all permutations of the pattern arguments.
However, with this naive approach, a total of $n!m!$ combinations have to be matched where
$n$ is the number of subject arguments and $m$ the number of pattern arguments.
It is likely that most of these combinations do not match or yield redundant matches.

Instead, we interpret the arguments as a multiset, i.e. an orderless collection that allows repetition of elements.
Also, we use the following order for matching the subterms of a commutative term:\newcounter{listcnt0}
\begin{list}{\arabic{listcnt0}.}
{
\usecounter{listcnt0}
\setlength{\rightmargin}{\leftmargin}
}

\item 

Constant arguments
\item 

Matched variables, i.e. variables that already have a value assigned in the current substitution
\item 

Non-variable arguments
\item 

Repeat step 2
\item 

Regular variables
\item 

Sequence variables\end{list}

Each of those steps reduces the search space for successive steps.
This also means that if one step finds no match, the remaining steps do not have to be performed.
Note that steps 3, 5 and 6 can yield multiple matches and backtracking is employed to check every combination.
Since step 6 is the most involved, it is described in more detail in the next section.

\subsubsection{Sequence Variables in Commutative Functions%
  \label{sequence-variables-in-commutative-functions}%
}

The distribution of $n$ subjects subterms onto $m$ sequence variables within a
commutative function symbol can yield up to $m^n$ distinct solutions.
Enumerating all of the solutions is accomplished by generating and solving several linear Diophantine equations.
As an example, lets assume we want to match \texttt{f(a, b, b, b)} with \texttt{f(x\_\_\_, y\_\_, y\_\_)} where \texttt{f} is commutative.
This means that the possible distributions are given by the non-negative integer solutions of these equations:\begin{eqnarray*}
1 &=& x_a + 2 y_a \\
3 &=& x_b + 2 y_b
\end{eqnarray*}$x_a$ determines how many times \texttt{a} is included in the substitution for \texttt{x}.
Because \texttt{y\_\_} requires at least one term, we have the additional constraint $y_a + y_b \geq 1$.
The only possible solution $x_a = x_b = y_b = 1 \wedge y_a = 0$ corresponds to the match substitution $\{ x \mapsto (a, b), y \mapsto (b) \}$.

Extensive research has been done on solving linear Diophantine equations and linear Diophantine
equation systems \DUrole{cite}{Weinstock1960,Bond1967,Lambert1988,Clausen1989,Aardal2000}.
In our case the equations are actually independent expect for the additional constraints for plus variables.
Also, the non-negative solutions can be found more easily.
We use an adaptation of the algorithm used in SymPy which recursively reduces any linear Diophantine equation to equations of the form $ax + by = d$.
Those can be solved efficiently with the Extended Euclidian algorithm \DUrole{cite}{Menezes1996}.
Then the solutions for those can be combined into a solution for the original equation.

All coefficients in those equations are likely very small since they correspond to the multiplicity of sequence variables.
Similarly, the number of variables in the equations is usually small as they map to sequence variables.
The constant is the multiplicity of a subject term and hence also usually small.
Overall, the number of distinct equations that are solved is small and the solutions are cached.
This reduces the impact of the sequence variables on the overall run time.

\subsection{Optimizations%
  \label{optimizations}%
}

Since most applications for pattern matching repeatedly match a fixed set of patterns against
multiple subjects, we implemented many-to-one matching for MatchPy.
The goal of many-to-one matching is to utilize similarities between patterns to match them more efficiently.
In this section, we give a brief overview of the many-to-one matching algorithm used by MatchPy.
Full details can be found in the master thesis \DUrole{cite}{thesis}.

\subsubsection{Many-to-one Matching%
  \label{many-to-one-matching}%
}

MatchPy includes two additional algorithms for matching: \texttt{ManyToOneMatcher} and \texttt{DiscriminationNet}.
Both enable matching multiple patterns against a single subject much faster than matching each pattern individually using \texttt{match}.
The latter can only be used for syntactic patterns and implements a state-of-the-art deterministic discrimination net.
A discrimination net is a data structure similar to a decision tree or a finite automaton \DUrole{cite}{Christian1993,Graef1991,Nedjah1997}.
The \texttt{ManyToOneMatcher} utilizes a generalized form of non-deterministic discrimination nets that support sequence variables and associative function symbols.
Furthermore, as elaborated in the next section, it can also match commutative terms.\begin{figure}[]\noindent\makebox[\columnwidth][c]{\includegraphics[width=\columnwidth]{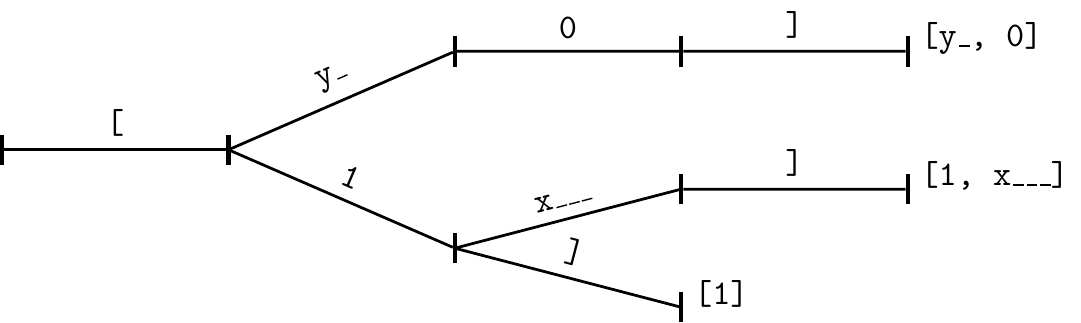}}
\caption{Example Discrimination Net. \DUrole{label}{fig:dn}}
\end{figure}

In Figure \DUrole{ref}{fig:dn}, an example for a non-deterministic discrimination net is shown.
It contains three patterns that match Python lists:
One matches the list that consists of a single 1, the second one matches a list with exactly two elements
where the last element is 0, and the third pattern matches any list where the first element is 1.
Note, that these patterns can also match nested lists, e.g. the second pattern would also match \texttt{{[}{[}2, 1{]}, 0{]}}.

Matching starts at the root and proceeds along the transitions.
Simultaneously, the subject is traversed in preorder and each symbol is checked against the transitions.
Only transitions matching the current subterm can be used.
Once a final state is reached, its label gives a list of matching patterns.
For non-deterministic discrimination nets, all possibilities need to be explored via backtracking.
The discrimination net allows to reduce the matching costs, because common parts of different pattern only need to be matched once.
For non-matching transitions, their whole subtree is pruned and all the patterns are excluded at once, further reducing the match cost.

In Figure \DUrole{ref}{fig:dn}, for the subject \texttt{{[}1, 0{]}}, there are two paths and therefore two matching patterns:
\texttt{{[}y\_, 0{]}} matches with $\{ y \mapsto 1 \}$ and \texttt{{[}1, x\_\_\_{]}} matches with $\{ x \mapsto 0 \}$.
Both the \texttt{y}-transition and the \texttt{1}-transition can be used in the second state to match a \texttt{1}.

Compared to existing discrimination net variants, we added transitions for the end of a compound term to support variadic functions.
Furthermore, we added support for both associative function symbols and sequence variables.
Finally, our discrimination net supports transitions restricted to symbol classes (i.e. \texttt{Symbol} subclasses)
in addition to the ones that match just a specific symbol.
We decided to use a non-deterministic discrimination net instead of a deterministic one, since the number
of states of the later would grow exponentially with the number of patterns.
While the \texttt{DiscriminationNet} also has support for sequence variables, in practice the net became to large to use with just a dozen patterns.

\subsubsection{Commutative Many-to-one Matching%
  \label{commutative-many-to-one-matching}%
}

Many-to-one matching for commutative terms is more involved.
We use a nested \texttt{CommutativeMatcher} which in turn uses another \texttt{ManyToOneMatcher} to match the subterms.
Our approach is similar to the one used by Bachmair and Kirchner in their respective works \DUrole{cite}{Bachmair1995,Kirchner2001}.
We match all the subterms of the commutative function in the subject with a many-to-one matcher constructed from the
subpatterns of the commutative function in the pattern (except for sequence variables, which are handled separately).
The resulting matches form a bipartite graph, where one set of nodes consists of the subject subterms and the other contains all the pattern subterms.
Two nodes are connected by an edge iff the pattern matches the subject.
Such an edge is also labeled with the match substitution(s).
Finding an overall match is then accomplished by finding a maximum matching in this graph.
However, for the matching to be valid, all the substitutions on its edges must be compatible,
i.e. they cannot have contradicting replacements for the same variable.
We use the Hopcroft-Karp algorithm \DUrole{cite}{Hopcroft1973} to find an initial maximum matching.
However, since we are also interested in all matches and the initial matching might have incompatible substitutions,
we use the algorithm described by Uno, Fukuda and Matsui \DUrole{cite}{Fukuda1994,Uno1997} to enumerate all maximum matchings.

To avoid yielding redundant matches, we extended the bipartite graph by introducing a total order over its two node sets.
This enables determining whether the edges of a matching maintain the order induced by the subjects or whether some of the edges \textquotedbl{}cross\textquotedbl{}.
Formally, for all edge pairs $(p, s), (p', s') \in M$ we require $(s \equiv s' \wedge p > p') \implies s > s'$
to hold where $M$ is the matching, $s, s'$ are subjects, and $p, p'$ are patterns.
An example of this is given in Figure \DUrole{ref}{fig:bipartite2}.
The order of the nodes is indicated by the numbers next to them.
The only two maximum matchings for this particular match graph are displayed.
In the left matching, the edges with the same subject cross and hence this matching is discarded.
The other matching is used because it maintains the order.
This ensures that only unique matches are yielded.
Once a matching for the subpatterns is obtained, the remaining subject arguments are
distributed to sequence variables in the same way as for one-to-one matching.\begin{figure}[]\noindent\makebox[\columnwidth][c]{\includegraphics[width=\columnwidth]{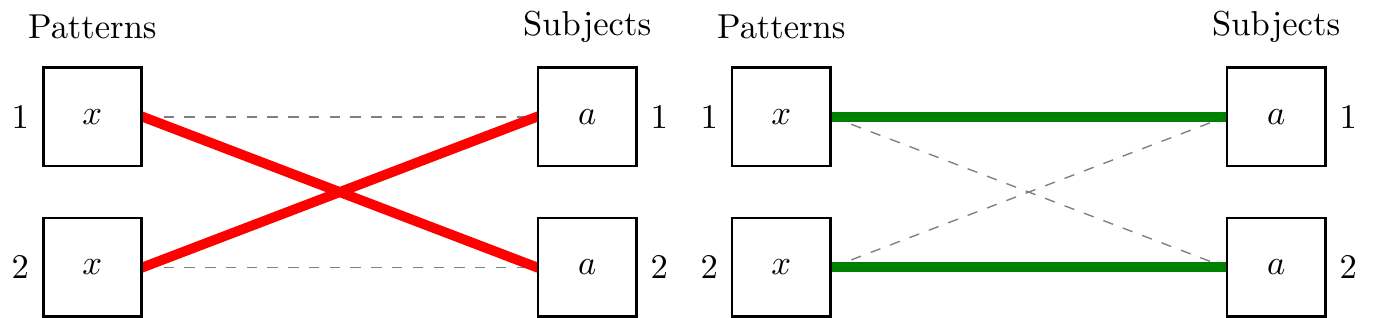}}
\caption{Example for Order in Bipartite Graph. \DUrole{label}{fig:bipartite2}}
\end{figure}

\subsection{Experiments%
  \label{experiments}%
}

To evaluate the performance of MatchPy, we conducted experiments on an Intel Core i5-2500K 3.3 GHz CPU with 8GB of RAM.
Our focus is on relative performance of one-to-one and many-to-one matching rather than the absolute performance.

\subsubsection{Linear Algebra%
  \label{linear-algebra}%
}

The operations for the linear algebra problem are shown in Table \DUrole{ref}{tbl:laop}.
The patterns all match BLAS kernels similar to the example pattern which was previously described.
The pattern set consists of 199 such patterns.
Out of those, 61 have an addition as outermost operation, 135 are patterns for products, and 3 are patterns for single matrices.
A lot of these patterns only differ in terms of constraints, e.g. there are ten distinct patterns
matching $A \times B$ with different constraints on the two matrices.
By removing the sequence variables from the product patterns, these pattern can be made syntactic when ignoring the multiplication's associativity.
In the following, we refer to the set of patterns with sequence variables as \texttt{LinAlg} and the set of syntactic product patterns as \texttt{Syntactic}.

The subjects were randomly generated such that matrices had random properties and each factor could randomly be transposed/inverted.
The number of factors was chosen according to a normal distribution with $\mu = 5$.
The total subject set consisted of 70 random products and 30 random sums.
Out of the pattern set, random subsets were used to examine the influence of the pattern set size on the matching time.
Across multiple subsets and repetitions per subject, the mean match and setup times were measured.
Matching was performed both with the \texttt{match} function and the \texttt{ManyToOneMatcher} (MTOM).
The results are displayed in Figure \DUrole{ref}{fig:linalgtime}.\begin{figure}[]\noindent\makebox[\columnwidth][c]{\includegraphics[width=\columnwidth]{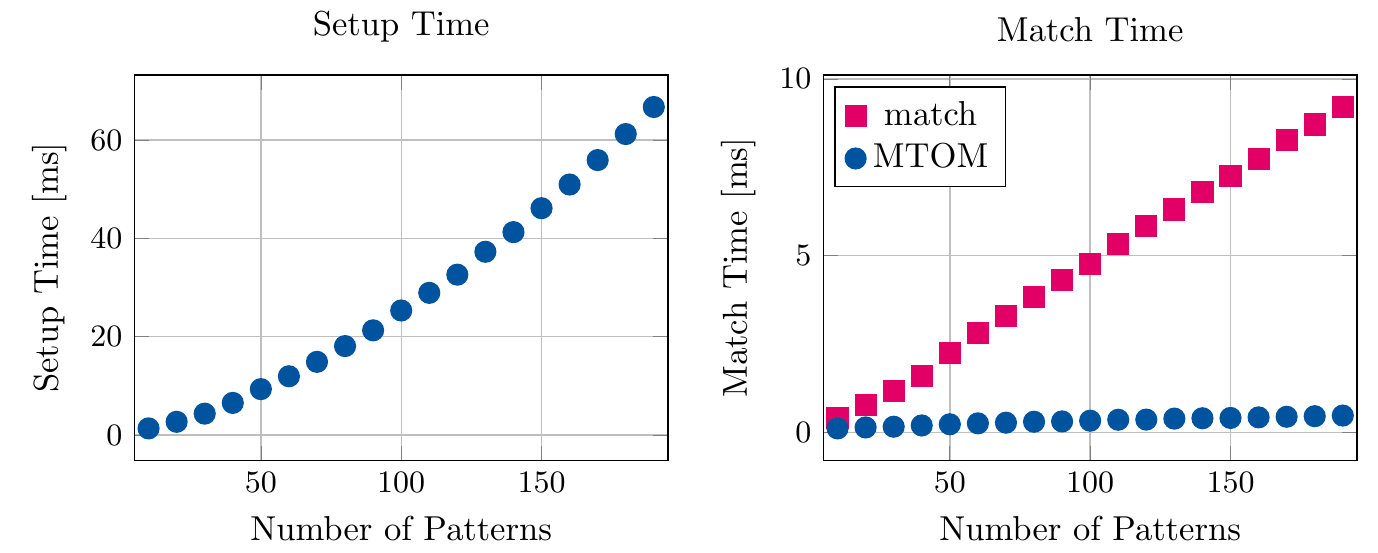}}
\caption{Timing Results for \texttt{LinAlg}. \DUrole{label}{fig:linalgtime}}
\end{figure}

As expected, both setup and match times grow with the pattern set size.
The growth of the many-to-one match time is much slower than the one for one-to-one matching.
This is also expected since the simultaneous matching is more efficient.
However, the growth of setup time for the many-to-one matcher beckons the question whether the speedup of the many-to-one matching is worth it.\begin{figure}[]\noindent\makebox[\columnwidth][c]{\includegraphics[width=\columnwidth]{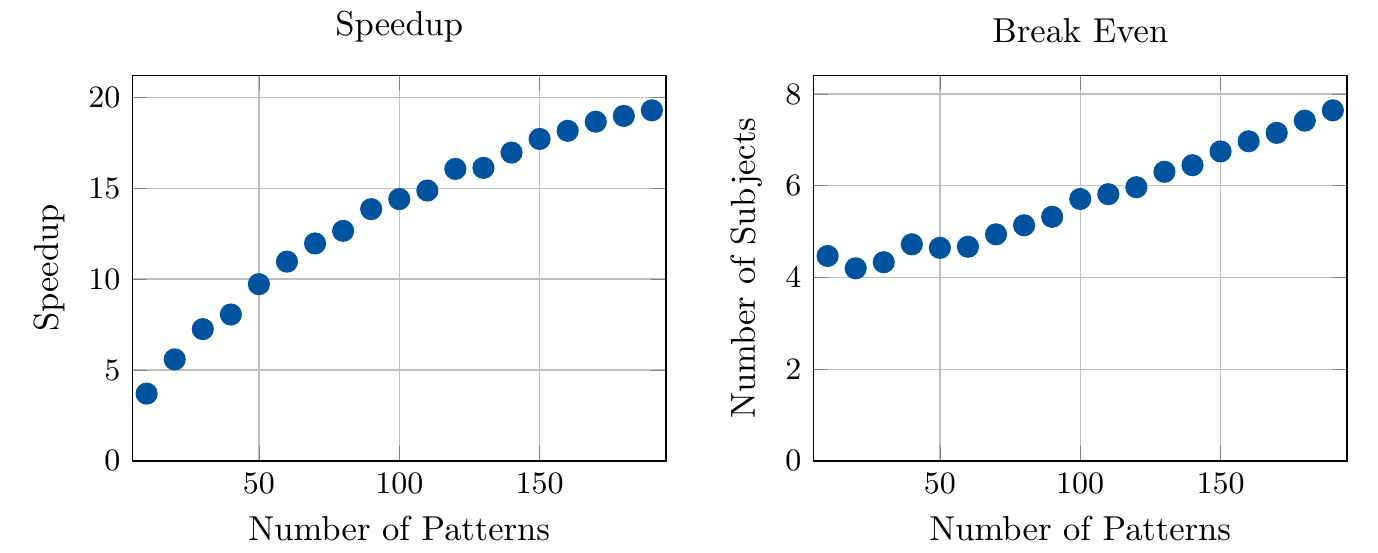}}
\caption{Comparison for \texttt{LinAlg}. \DUrole{label}{fig:linalgspeed}}
\end{figure}

Figure \DUrole{ref}{fig:linalgspeed} depicts both the speedup and the break even point for many-to-one matching for \texttt{LinAlg}.
The first graph indicates that the speedup of many-to-one matching increases with larger pattern sets.
But in order to profit from that speedup, the setup cost of many-to-one matching must be amortized.
Therefore, the second graph shows the break even point for many-to-one matching in terms of number of subjects.
If for a given number of patterns and subjects the corresponding point is above the line, then many-to-one matching is overall faster.
In this example, when matching more than eight times, many-to-one matching is overall always faster than one-to-one matching.

For the syntactic product patterns we compared the \texttt{match} function, the \texttt{ManyToOneMatcher} (MTOM) and the \texttt{DiscriminationNet} (DN).
Again, randomly generated subjects were used.
The resulting speedups and break even points are displayed in Figure \DUrole{ref}{fig:syntacticspeed}.\begin{figure}[]\noindent\makebox[\columnwidth][c]{\includegraphics[width=\columnwidth]{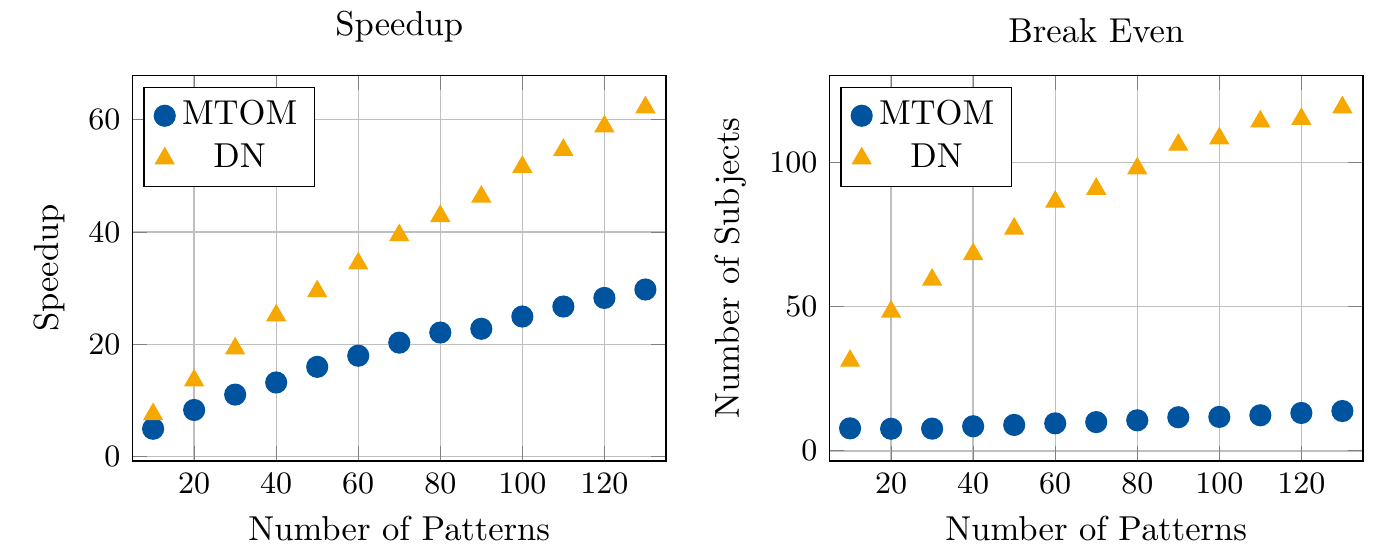}}
\caption{Comparison for \texttt{Syntactic}. \DUrole{label}{fig:syntacticspeed}}
\end{figure}

In this case, the discrimination net is the fastest overall reaching a speedup of up to 60.
However, because it also has the highest setup time, it only outperforms the many-to-one matcher after about 100 subjects for larger pattern set sizes.
In practice, the discrimination net is likely the best choice for syntactic patterns, as long as the discrimination net does not grow to large.
In the worst case, the size of the discrimination net can grow exponentially in the number of patterns.

\subsubsection{Abstract Syntax Trees%
  \label{abstract-syntax-trees}%
}

Python includes a tool to convert code from Python 2 to Python 3.
It is part of the standard library package \texttt{lib2to3} which has a collection of \textquotedbl{}fixers\textquotedbl{} that each convert one of the incompatible cases.
To find matching parts of the code, those fixers use pattern matching on the abstract syntax tree (AST).
Such an AST can be represented in the MatchPy data structures.
We converted some of the patterns used by \texttt{lib2to3} both to demonstrate the generality of MatchPy and to evaluate the performance of many-to-one matching.
Because the fixers are applied one after another and can modify the AST after each match,
it would be difficult to use many-to-one matching for \texttt{lib2to3} in practice.

The following is an example of such a pattern:\vspace{1mm}
\begin{Verbatim}[commandchars=\\\{\},fontsize=\footnotesize]
\PY{n}{power}\PY{o}{\PYZlt{}}
    \PY{l+s+s1}{\PYZsq{}}\PY{l+s+s1}{isinstance}\PY{l+s+s1}{\PYZsq{}}
    \PY{n}{trailer}\PY{o}{\PYZlt{}} \PY{l+s+s1}{\PYZsq{}}\PY{l+s+s1}{(}\PY{l+s+s1}{\PYZsq{}} \PY{n}{arglist}\PY{o}{\PYZlt{}} \PY{n+nb}{any} \PY{l+s+s1}{\PYZsq{}}\PY{l+s+s1}{,}\PY{l+s+s1}{\PYZsq{}} \PY{n}{atom}\PY{o}{\PYZlt{}} \PY{l+s+s1}{\PYZsq{}}\PY{l+s+s1}{(}\PY{l+s+s1}{\PYZsq{}}
        \PY{n}{args}\PY{o}{=}\PY{n}{testlist\PYZus{}gexp}\PY{o}{\PYZlt{}} \PY{n+nb}{any}\PY{o}{+} \PY{o}{\PYZgt{}}
    \PY{l+s+s1}{\PYZsq{}}\PY{l+s+s1}{)}\PY{l+s+s1}{\PYZsq{}} \PY{o}{\PYZgt{}} \PY{o}{\PYZgt{}} \PY{l+s+s1}{\PYZsq{}}\PY{l+s+s1}{)}\PY{l+s+s1}{\PYZsq{}} \PY{o}{\PYZgt{}}
\PY{o}{\PYZgt{}}
\end{Verbatim}
\vspace{1mm}
It matches an \texttt{isinstance} expression with a tuple as second argument.
Its tree structure is illustrated in Figure \DUrole{ref}{fig:ast}.
The corresponding fixer cleans up duplications generated by previous fixers.
For example \texttt{\DUrole{code}{\DUrole{py}{\DUrole{python}{\DUrole{name}{\DUrole{builtin}{isinstance}}\DUrole{punctuation}{(}\DUrole{name}{x}\DUrole{punctuation}{,} \DUrole{punctuation}{(}\DUrole{name}{\DUrole{builtin}{int}}\DUrole{punctuation}{,} \DUrole{name}{\DUrole{builtin}{long}}\DUrole{punctuation}{))}}}}} would be converted by another fixer into \texttt{\DUrole{code}{\DUrole{py}{\DUrole{python}{\DUrole{name}{\DUrole{builtin}{isinstance}}\DUrole{punctuation}{(}\DUrole{name}{x}\DUrole{punctuation}{,} \DUrole{punctuation}{(}\DUrole{name}{\DUrole{builtin}{int}}\DUrole{punctuation}{,} \DUrole{name}{\DUrole{builtin}{int}}\DUrole{punctuation}{))}}}}},
which in turn is then simplified to \texttt{\DUrole{code}{\DUrole{py}{\DUrole{python}{\DUrole{name}{\DUrole{builtin}{isinstance}}\DUrole{punctuation}{(}\DUrole{name}{x}\DUrole{punctuation}{,} \DUrole{name}{\DUrole{builtin}{int}}\DUrole{punctuation}{)}}}}} by this fixer.\begin{figure}[]\noindent\makebox[\columnwidth][c]{\includegraphics[scale=0.80]{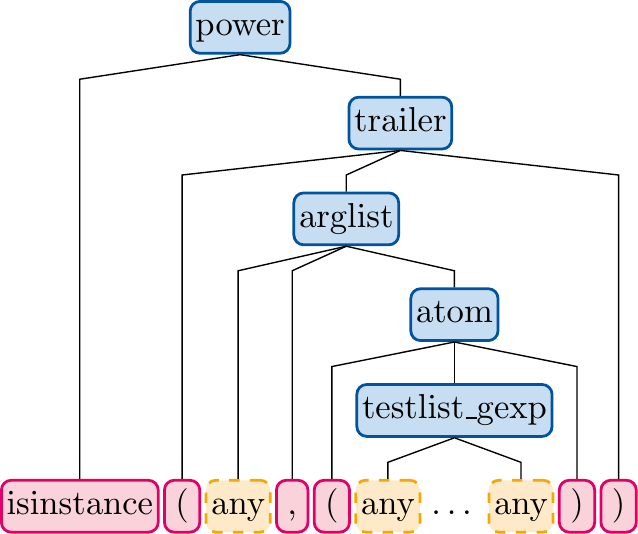}}
\caption{AST of the \texttt{isinstance} pattern. \DUrole{label}{fig:ast}}
\end{figure}

Out of the original 46 patterns, 36 could be converted to MatchPy patterns.
Some patterns could not be converted, because they contain features that MatchPy does not support yet.
The features include negated subpatterns (e.g. \texttt{\DUrole{code}{\DUrole{py}{\DUrole{python}{\DUrole{operator}{\DUrole{word}{not}} \DUrole{name}{atom}\DUrole{operator}{<}\DUrole{literal}{\DUrole{string}{\DUrole{single}{'('}}} \DUrole{punctuation}{{[}}\DUrole{name}{\DUrole{builtin}{any}}\DUrole{punctuation}{{]}} \DUrole{literal}{\DUrole{string}{\DUrole{single}{')'}}}\DUrole{operator}{>}}}}})
and subpatterns that allow an aritrary number of repetitions (e.g. \texttt{\DUrole{code}{\DUrole{py}{\DUrole{python}{\DUrole{name}{\DUrole{builtin}{any}} \DUrole{punctuation}{(}\DUrole{literal}{\DUrole{string}{\DUrole{single}{','}}} \DUrole{name}{\DUrole{builtin}{any}}\DUrole{punctuation}{)}\DUrole{operator}{+}}}}}).

Furthermore, some of the AST patterns contain alternative or optional subpatterns, e.g. \texttt{\DUrole{code}{\DUrole{py}{\DUrole{python}{\DUrole{name}{power}\DUrole{operator}{<}\DUrole{literal}{\DUrole{string}{\DUrole{single}{'input'}}} \DUrole{name}{args}\DUrole{operator}{=}\DUrole{name}{trailer}\DUrole{operator}{<}\DUrole{literal}{\DUrole{string}{\DUrole{single}{'('}}} \DUrole{punctuation}{{[}}\DUrole{name}{\DUrole{builtin}{any}}\DUrole{punctuation}{{]}} \DUrole{literal}{\DUrole{string}{\DUrole{single}{')'}}}\DUrole{operator}{>{}>}}}}}.
These features are also not directly supported by MatchPy, but they can be replicated by using multiple patterns.
For those \texttt{lib2to3} patterns, all combinations of the alternatives were generated and added as invividual patterns.
This resulted in about 1200 patterns for the many-to-one matcher that completely cover the original 36 patterns.

For the experiments, we used a file that combines the examples from the unittests of \texttt{lib2to3} with about 900 non-empty lines.
We compared the set of 36 patterns with the original matcher and the 1200 patterns with the many-to-one matcher.
A total of about 560 matches are found.
Overall, on average, our many-to-one matcher takes 0.7 seconds to find all matches, while the matcher from \texttt{lib2to3} takes 1.8 seconds.
This yields a speedup of approximately 2.5. However, the construction of the many-to-one matcher takes 1.4 seconds on average.
However, this setup cost will be amortized by the faster matching for sufficiently large ASTs.
The setup time can also mostly be eliminated by saving the many-to-one matcher to disk and loading it once required.

Compared to the one-to-one matching in MatchPy, the many-to-one matching achieves a speedup of about 60.
This is due to the fact that for any given subject less than 1\% of patterns match.
By taking into account the setup time of the many-to-one matcher, the break even point for it is at about 200 subjects.

\subsection{Conclusions%
  \label{conclusions}%
}

We have presented MatchPy, a pattern matching library for Python with support for sequence variables and associative/commutative functions.
This library includes algorithms and data structures for both one-to-one and many-to-one matching.
Because non-syntactic pattern matching is NP-hard, in the worst case the pattern matching times grows exponentially with the length of the pattern.
Nonetheless, our experiments on real world examples indicate that many-to-one matching can give a significant speedup over one-to-one matching.
However, the employed discrimination nets come with a one-time construction cost which needs to be amortized to benefit from their speedup.
In our experiments, the break even point for many-to-one matching was always reached well within the typical number of subjects for the respective application.
Therefore, many-to-one matching is likely to result in a compelling speedup in practice.

For syntactic patterns, we also compared the syntactic discrimination net with the many-to-one matcher.
As expected, discrimination nets are faster at matching, but also have a significantly higher setup time.
Furthermore, the number of states can grow exponentially with the number of patterns, making them unsuitable for some pattern sets.
Overall, if applicable, discrimination nets offer better performance than a many-to-one matcher.

Which pattern matching algorithm is the fastest for a given application depends on many factors.
Hence, it is not possible to give a general recommendation.
Yet, the more subjects are matched against the same pattern set, the more likely it is that many-to-one outperforms one-to-one matching.
In the experiments, a higher number of patterns lead to an increase of the speedup of many-to-one matching.
In terms of the size of the many-to-one matcher, the growth of the net was sublinear in our experiments and still feasible for large pattern sets.
The efficiency of using many-to-one matching also heavily depends on the actual pattern set, i.e. the degree of similarity and overlap between the patterns.

\subsection{Future Work%
  \label{future-work}%
}

We plan on extending MatchPy with more powerful pattern matching features to make it useful for an even wider range of applications.
The greatest challenge with additional features is likely to implement them for many-to-one matching.
In the following, we discuss some possibilities for extending the library.

\subsubsection{Additional pattern features%
  \label{additional-pattern-features}%
}

In the future, we plan to implement similar functionality to the \texttt{Repeated}, \texttt{Sequence}, and \texttt{Alternatives} functions from Mathematica.
These provide another level of expressive power which cannot be fully replicated with the current feature set of MatchPy.
Another useful feature are context variables as described by Kutsia \DUrole{cite}{Kutsia2006}.
They allow matching subterms at arbitrary depths which is especially useful for structures like XML.
With context variables, MatchPy's pattern matching would be as powerful as XPath \DUrole{cite}{Robie2017} or CSS selectors \DUrole{cite}{Rivoal2017} for such structures.
Similarly, function variables which can match a function symbol would also be useful for those applications.

\subsubsection{Integration%
  \label{integration}%
}

Currently, in order to use MatchPy, existing data structures must be adapted to provide their children via an iterator.
Where that is not possible, for example because the data structures are provided by a third party library, translation functions need to be applied.
Also, some native data structures such as dicts are currently not supported directly.
Therefore, it would be useful, to have a better way of using existing data structures with MatchPy.

In particular, easy integration with SymPy is an important goal, since it is a popular tool for working with symbolic mathematics.
SymPy already implements \href{http://docs.sympy.org/0.7.2/tutorial.html\#pattern-matching}{a form of pattern matching} which is less powerful than MatchPy.
It lacks support for sequence variables, symbol wildcards and constraints.
Each constant symbol in SymPy can have properties that allow it to be commutative or non-commutative.
One benefit of this approach is easier modeling of linear algebra multiplication, where matrices and vectors do not commute, but scalars do.
Better integration of MatchPy with SymPy would provide the users of SymPy with more powerful pattern matching tools.
However, Matchpy would require selective commutativity to be fully compatible with SymPy.
Also, SymPy supports older Python versions, while MatchPy requires Python 3.6.

\subsubsection{Performance%
  \label{performance}%
}

If pattern matching is a major part of an application, its running time can significantly impact the overall speed.
Reimplementing parts of MatchPy as a C module would likely result in a substantial speedup.
Alternatively, adapting part of the code to Cython could be another option to increase the speed \DUrole{cite}{Behnel2009, Wilbers2009}.
Furthermore, generating source code for a pattern set similar to parser generators for formal grammars could improve matching performance.
While code generation for syntactic pattern matching has been the subject of various works \DUrole{cite}{Augustsson1985,Fessant2001,Maranget2008,Moreau2003},
its application with the extended feature set of MatchPy is another potential area of future research.
Also, additonal research on the viability of pattern matching with increasingly complex and large subjects or patterns is desirable.
Parallelizing many-to-one matching is also a possibility to increase the overall speed which is worth exploring.

\subsubsection{Functional pattern matching%
  \label{functional-pattern-matching}%
}

Since Python does not have pattern matching as a language feature, MatchPy could be extended to provide a syntax similar to other functional programming languages.
However, without a switch statement as part of the language, there is a limit to the syntax of this pattern expression.
The following is an example of what such a syntax could look like:\vspace{1mm}
\begin{Verbatim}[commandchars=\\\{\},fontsize=\footnotesize]
\PY{k}{with} \PY{n}{match}\PY{p}{(}\PY{n}{f}\PY{p}{(}\PY{n}{a}\PY{p}{,} \PY{n}{b}\PY{p}{)}\PY{p}{)}\PY{p}{:}
    \PY{k}{if} \PY{n}{case}\PY{p}{(}\PY{n}{f}\PY{p}{(}\PY{n}{x\PYZus{}}\PY{p}{,} \PY{n}{y\PYZus{}}\PY{p}{)}\PY{p}{)}\PY{p}{:}
        \PY{k}{print}\PY{p}{(}\PY{l+s+s2}{\PYZdq{}}\PY{l+s+s2}{x=\PYZob{}\PYZcb{}, y=\PYZob{}\PYZcb{}}\PY{l+s+s2}{\PYZdq{}}\PY{o}{.}\PY{n}{format}\PY{p}{(}\PY{n}{x}\PY{p}{,} \PY{n}{y}\PY{p}{)}\PY{p}{)}\PY{p}{)}
    \PY{k}{elif} \PY{n}{case}\PY{p}{(}\PY{n}{f}\PY{p}{(}\PY{n}{z\PYZus{}}\PY{p}{)}\PY{p}{)}\PY{p}{:}
        \PY{o}{.}\PY{o}{.}\PY{o}{.}\PY{o}{.}
\end{Verbatim}
\vspace{1mm}
There are already several libraries for Python which implement such a functionality for syntactic
patterns and native data structures (e.g. MacroPy \DUrole{cite}{MacroPy}, patterns \DUrole{cite}{patterns} or PyPatt \DUrole{cite}{PyPatt}).
However, the usefulness of this feature needs further evaluation.
\bibliographystyle{alphaurl}
\bibliography{literature}

\end{document}